\documentclass[referee]{aa}

\usepackage{amsmath}
\usepackage{amssymb}
\usepackage{latexsym}
\usepackage{subfig,graphicx}
\usepackage{txfonts}
\usepackage{natbib}
\bibpunct{(}{)}{,}{}{}{,}
\usepackage{color}
\usepackage[mathcal]{euscript} %Allows order of symbol \mathcal{O}

\begin{document}
\title{Non-axisymmetric oscillations of stratified coronal magnetic loops
with elliptical cross-sections}

\author{ R. J. Morton and M. S. Ruderman}

\institute{Solar Physics and Space Plasma Research Centre
(SP$^2$RC), University of Sheffield, Hicks Building, Hounsfield
Road, Sheffield S3 7RH, UK
\\email:[r.j.morton; m.s.ruderman]@sheffield.ac.uk}

\date{Received /Accepted}

\abstract{}{We study non-axisymmetric oscillations of a straight
magnetic tube with an elliptic cross-section and density varying
along the tube.} {The governing equations for kink and fluting modes
in the thin tube approximation are derived. We found that there are
two kink modes, polarised along the large and small axes of the
elliptic cross-section. We have shown that the ratio of frequencies
of the first overtone and fundamental harmonic is the same for both
kink modes and independent of the ratio of the ellipse axes.}{ On
the basis of this result we concluded that the estimates of the
atmospheric scale height obtained using simultaneous observations of
the fundamental harmonic and first overtone of the coronal loop kink
oscillations are independent of the ellipticity of the loop
cross-section.}{}


\titlerunning{Oscillations of loops with elliptical cross-sections}
\authorrunning{R.J Morton and M. S. Ruderman}

\maketitle

\section{Introduction}
\label{sec:intro}
The solar atmosphere is a highly dynamic and
structured plasma that is able to support a wide variety of
magneto-acoustic waves and oscillations. Each layer of the solar
atmosphere, from the photosphere to the corona, is magnetically
connected to the others via the all pervading magnetic field. The
omnipresence of the waves throughout the atmosphere is becoming well
documented as new and exciting techniques are being developed to
help observe and study the waves (see, e.g. \citealp{BANETAL2007};
\citealp{TOMetal2007}).

After transverse coronal loop oscillations were first observed by
TRACE in $1998$ (\citealp{ASCetal1999}; \citealp{NAKetal1999}), the
phenomenon became one of the hot topics within solar physics. In the
first theoretical interpretation of these oscillations, a coronal
loop was modelled as a straight magnetic cylinder with the density
constant inside and outside. Since then, a number of more
complicated and realistic models have been considered. For a recent
review on the theory of transverse oscillations of a coronal loop
see, e.g., \cite{RUDERD2009}.

Although the transverse coronal loop oscillations are interesting on
their own, their main importance is related to the fact that they
are a powerful tool of coronal seismology. \cite{NAKOFM2001}
demonstrated this by using the observations of transverse coronal
loop oscillations to estimate the magnitude of the magnetic field
in the corona, while \cite{ANDetal2005a} suggested to use these
observations to estimate the atmospheric scale height in the corona.

In this paper we continue to study the transverse oscillations of
coronal loops. Coronal loops with elliptical cross-sections and a
\emph{constant} density profile have been studied previously in both
cold (\citealp{RUD2003}) and { finite-}\/$\beta$
(\citealp{ERDMOR2009}) plasmas. Now, we consider oscillations of
loops with the density \emph{varying} along the loop and a constant
elliptic cross-section. The paper is organized as follows.  In the
next section we formulate the problem. In Sect.~\ref{sec:derivation}
we derive the governing equations for non-axisymmetric oscillations
of a coronal loop with an elliptic cross-section in the thin tube
approximation. In Sect.~\ref{sec:seismology} we study the
implication of our analysis on coronal seismology.
Section~\ref{sec:summary} contains the summary of the obtained
results and our conclusions.

\section{Problem formulation}
\label{sec:formulation}

We model a coronal loop as a straight magnetic tube with an
elliptical cross-section. The cold plasma approximation is used. The
density varies along the tube, while the cross-section remains
constant. In Cartesian coordinates $x,\,y,\,z$ the loop axis
coincides with the $z$\/-axis. The equilibrium magnetic field is
given by $\vec{B} = B\hat{\vec{z}}$\/, where $B$ is constant and
$\hat{\vec{z}}$ is the unit vector in the $z$\/-direction. The
plasma motion is governed by the linearised ideal MHD equations,
\begin{equation}
\frac{\partial^2\vec{\xi}}{\partial t^2} =
   \frac1{\mu_0\rho}(\nabla\times\vec{b})\times\vec{B},
\label{eq:ideal_v}
\end{equation}
\begin{equation}
\vec{b} = \nabla\times(\vec{\xi}\times\vec{B}).
\label{eq:ideal_b}
\end{equation}
Here $\vec{\xi}$ is the plasma displacement, $\vec{b}$ the magnetic field
perturbation, $\rho(z)$ the equilibrium density, and $\mu_0$ the magnetic
permeability of free space; $\rho(z) = \rho_{\rm i}(z)$ inside the tube and
$\rho(z) = \rho_{\rm e}(z)$ outside the tube.

\begin{figure}
\centering
\includegraphics[scale=0.7]{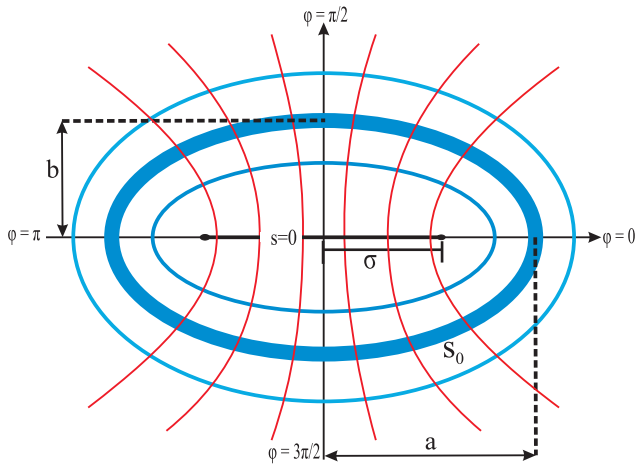}
\caption{Sketch showing the elliptical coordinate system used to
describe the loop cross-section. { The open and closed curves show
the $s$ and $\varphi$ coordinate lines respectively. The thick
closed curve shows the tube boundary.}}\label{fig:1}
\end{figure}

Let us introduce the elliptic coordinates $s$ and $\varphi$ in the
$xy$\/-plane (see Fig.~\ref{fig:1}). The Cartesian coordinates are
expressed in terms of elliptic coordinates as
\begin{equation}
x = \sigma\cosh s\cos\varphi, \qquad y = \sigma\sinh s\sin\varphi,
\label{eq:ellip_coord}
\end{equation}
where $\sigma$ is a quantity with the dimension of length, $s$ varies from $0$
to $\infty$\/, and $\varphi$ from $-\pi$ to $\pi$\/. In the elliptic
coordinates the equation of the tube boundary is $s = s_0$\/. Then the large
and small half-axes of the tube elliptic cross-section are in the $x$ and
$y$\/-direction, and they are given by
\begin{equation}
a = \sigma\cosh s_0, \qquad b = \sigma\sinh s_0 .
\label{eq:axes}
\end{equation}
At the tube boundary the normal component of the displacement, $\xi_s$\/, and
the magnetic pressure perturbation, $P = \vec{b}\cdot\vec{B}/\mu_0$\/, has to
be continuous,
\begin{equation}
[\hspace*{-0.7mm}[\xi_s]\hspace*{-0.7mm}] = 0, \quad
[\hspace*{-0.7mm}[P]\hspace*{-0.7mm}] = 0 \quad \mbox{at} \quad s = s_0,
\label{eq:jumps}
\end{equation}
where $[\hspace*{-0.7mm}[f]\hspace*{-0.7mm}]$ indicates the jumps of function
$f$ across the boundary defined as
\begin{equation}
[\hspace*{-0.7mm}[f]\hspace*{-0.7mm}] = \lim_{\varepsilon\to 0}
   [f(s + \varepsilon) - f(s - \varepsilon)].
\label{eq:jump-def}
\end{equation}

The magnetic field lines at the loop foot points are frozen in the dense
photospheric plasma, so that
\begin{equation}
\vec{\xi} = 0 \quad \mbox{at} \quad z = \pm L/2,
\label{eq:frozen-xi}
\end{equation}
where $L$ is the loop length.

It follows from Eq.~(\ref{eq:ellip_coord}) that the points with the elliptical
coordinates $s = 0$, $\varphi = \varphi_0$\/, and $s=0$, $\varphi = -\varphi_0$
are the same point in the $xy$\/-plane. This implies that $P$ and $\xi_s$ have
to satisfy the boundary conditions
\begin{equation}
P(0,\varphi) = P(0,-\varphi), \qquad \xi_s(0,\varphi) = -\xi_s(0,-\varphi).
\label{eq:reg_pxi}
\end{equation}
Equations (\ref{eq:ideal_v}) and (\ref{eq:ideal_b}) together with the boundary
conditions (\ref{eq:jumps}), (\ref{eq:frozen-xi}) and (\ref{eq:reg_pxi}) will be
used in the next section to derive the governing equations for non-axisymmetric
oscillations in the thin tube approximation.

\section{Derivation of governing equations}
\label{sec:derivation}

The analysis in this section is similar to one used by
\cite{DYMRUD2005} to derive the governing equation for a thin tube
with a circular tube cross-section. We begin by noting that, in
accordance with Eq.~(\ref{eq:ideal_v}), $\xi_z = 0$. The system of
Eqs.~(\ref{eq:ideal_v}) and (\ref{eq:ideal_b}) can then be
transformed to
\begin{equation}
\frac{\partial^2\vec{\xi}}{\partial t^2} = -\frac1\rho\nabla_\perp P +
   \frac B{\mu_0\rho}\frac{\partial\vec{b}_\perp}{\partial z},
\label{eq:transf_v}
\end{equation}
\begin{equation}
\vec{b}_\perp = B\frac{\partial\vec{\xi}}{\partial z},
\label{eq:transf_b}
\end{equation}
\begin{equation}
P = -\rho v_A^2\nabla\cdot\vec{\xi},
\label{eq:P2xi}
\end{equation}
where $v_A$ is the Alfv\'en speed defined by $v_A^2 = B^2/\mu_0\rho$\/, and the
operator $\nabla_\perp$ and component of the magnetic field perturbation
perpendicular to the $z$\/-axis are given by
\begin{equation}
\nabla_\perp = \nabla - \hat{\vec{z}}\frac\partial{\partial z}, \qquad
\vec{b}_\perp = \vec{b} - \vec{b}\cdot\hat{\vec{z}}.
\label{perp}
\end{equation}
Eliminating $\vec{b}_\perp$ from Eqs.~(\ref{eq:transf_v}) yields
\begin{equation}
\frac{\partial^2\vec{\xi}}{\partial t^2} -
   v_A^2\frac{\partial^2\vec{\xi}}{\partial z^2} = -\frac1\rho\nabla_\perp P.
\label{eq:momentum}
\end{equation}
Taking the divergence of this equation and using Eq.~(\ref{eq:P2xi}) we arrive
at the equation for $P$\/,
\begin{equation}
\frac{\partial^2 P}{\partial t^2} -
   v_A^2\frac{\partial^2 P}{\partial z^2} = v_A^2\nabla_\perp^2 P.
\label{eq:Pfull}
\end{equation}
Now we use the thin tube approximation. To do this we note that the
characteristic spatial scale in the $z$\/-direction is $L$\/, and
the characteristic time of the problem is $L/\bar{v}_A$\/, where
$\bar{v}_A$ is a typical value of Alfv\'en speed. In what follows we
only consider the perturbations that decay at the distance of a few
$a$ from the tube. Then the characteristic spatial scale in the $x$
and $y$\/-direction is $a$\/. It follows from this analysis that the
ratio of the left-hand side of Eq.~(\ref{eq:Pfull}) to its
right-hand side is of the order of $(a/L)^2 \ll 1$, so that we can
neglect the left-hand side. Then, using the expression for
$\nabla_\perp^2$ in the elliptical coordinates (e.g.
\citealp{KornKorn}), we obtain the equation for $P$ in the thin tube
approximation,
\begin{equation}
\frac{\partial^2 P}{\partial s^2} + \frac{\partial^2 P}{\partial\varphi^2} = 0.
\label{eq:Papprox}
\end{equation}
The solution to this equation has to satisfy the first regularity condition in
Eq.~(\ref{eq:reg_pxi}), and the second boundary condition in
Eq.~(\ref{eq:jumps}). Using Eq.~(\ref{eq:momentum}) we rewrite the second
regularity condition in terms of $P$\/,
\begin{equation}
\frac{\partial P(s,\varphi)}{\partial s}\bigg|_{s=0} =
   -\frac{\partial P(s,-\varphi)}{\partial s}\bigg|_{s=0}.
\label{eq:reg_derP}
\end{equation}

To derive the governing equations for non-axisymmetric tube oscillations we
solve Eqs.~(\ref{eq:momentum}) and (\ref{eq:Papprox}) inside and outside the
tube, and then match the two solutions at the tube boundary. It is
straightforward to obtain the general solution to Eq.~(\ref{eq:Papprox})
inside the tube satisfying the regularity conditions Eqs.~(\ref{eq:reg_pxi})
and (\ref{eq:reg_derP}),
\begin{equation}
P^{\rm i} = \sum_{n=1}^\infty\left[C_n^{\rm i}\cosh(ns)\cos(n\varphi) +
   D_n^{\rm i}\sinh(ns)\sin(n\varphi)\right],
\label{eq:p_int}
\end{equation}
where $C_n^{\rm i}$ and $D_n^{\rm i}$ are arbitrary functions of $t$ and $z$\/.
The solution outside the tube has to decay as $s \to \infty$\/. Hence, its
general form is
\begin{equation}
P^{\rm e} = \sum_{n=1}^\infty e^{-ns}\left[C_n^{\rm e}\cos(n\varphi) +
   D_n^{\rm e}\sin(n\varphi)\right],
\label{eq:p_ext}
\end{equation}
where once again $C_n^{\rm e}$ and $D_n^{\rm e}$ are arbitrary functions of $t$
and $z$\/. Substituting Eqs.~(\ref{eq:p_int}) and (\ref{eq:p_ext}) in
Eq.~(\ref{eq:momentum}) and using the expression for $\nabla_\perp$ in the
elliptical coordinates (e.g. \citealp{KornKorn}),
\begin{equation}
\nabla_\perp = \frac1{\sigma\Theta}
   \left(\hat{\vec{s}}\frac\partial{\partial s} +
   \hat{\vec{\varphi}}\frac\partial{\partial\varphi}\right), \quad
\Theta = (\sinh^2 s + \sin^2\varphi)^{1/2},
\label{eq:nabla2coord}
\end{equation}
where $\hat{\vec{s}}$ and $\hat{\vec{\varphi}}$ are the unit vectors in the $s$
and $\varphi$\/-direction, we obtain the expressions for $\xi_s$ inside and
outside the tube,
\begin{equation}
\xi_s^{\rm i} = \frac1{\sigma\Theta}\sum_{n=1}^\infty
   \left[F_n^{\rm i}\sinh(ns)\cos(n\varphi) +
   G_n^{\rm i}\cosh(ns)\sin(n\varphi)\right],
\label{eq:xi_int}
\end{equation}
\begin{equation}
\xi_s^{\rm e} = \frac1{\sigma\Theta}\sum_{n=1}^\infty e^{-ns}
   \left[F_n^{\rm e}\cos(n\varphi) + G_n^{\rm e}\sin(n\varphi)\right].
\label{eq:xi_ext}
\end{equation}
In these equations $F_n^{\rm i}$\/, $G_n^{\rm i}$\/, $F_n^{\rm e}$
and $G_n^{\rm e}$ are functions of $t$ and $z$\/. They are related
to the functions $C_n^{\rm i}$\/ $D_n^{\rm i}$\/, $C_n^{\rm e}$ and
$D_n^{\rm e}$ by
\begin{equation}
\frac{\partial^2 F_n^{\rm i}}{\partial t^2} -
   v_{A\rm i}^2\frac{\partial^2 F_n^{\rm i}}{\partial z^2} =
   -\frac{C_n^{\rm i}}{\rho_{\rm i}},
\label{eq:FCint}
\end{equation}
\begin{equation}
\frac{\partial^2 G_n^{\rm i}}{\partial t^2} -
   v_{A\rm i}^2\frac{\partial^2 G_n^{\rm i}}{\partial z^2} =
   -\frac{D_n^{\rm i}}{\rho_{\rm i}},
\label{eq:GDint}
\end{equation}
\begin{equation}
\frac{\partial^2 F_n^{\rm e}}{\partial t^2} -
   v_{A\rm e}^2\frac{\partial^2 F_n^{\rm e}}{\partial z^2} =
   \frac{C_n^{\rm e}}{\rho_{\rm e}},
\label{eq:FCext}
\end{equation}
\begin{equation}
\frac{\partial^2 G_n^{\rm e}}{\partial t^2} -
   v_{A\rm e}^2\frac{\partial^2 G_n^{\rm e}}{\partial z^2} =
   \frac{D_n^{\rm e}}{\rho_{\rm e}}.
\label{eq:GDext}
\end{equation}
Substituting Eqs.~(\ref{eq:p_int}) and (\ref{eq:p_ext}) in the second boundary
condition in Eq.~(\ref{eq:jumps}) we obtain
\begin{equation}
C_n^{\rm i}\cosh(ns_0) = e^{-ns_0}C_n^{\rm e}, \quad
D_n^{\rm i}\sinh(ns) = e^{-ns_0}D_n^{\rm e} .
\label{eq:CDint-ext}
\end{equation}
Substituting Eqs.~(\ref{eq:xi_int}) and (\ref{eq:xi_ext}) in the first boundary
condition in Eq.~(\ref{eq:jumps}) yields
\begin{equation}
F_n^{\rm i}\sinh(ns_0) = e^{-ns_0}F_n^{\rm e}, \quad
G_n^{\rm i}\cosh(ns) = e^{-ns_0}G_n^{\rm e} .
\label{eq:FGint-ext}
\end{equation}
Eliminating $C_n^{\rm i}$\/, $C_n^{\rm e}$ and $F_n^{\rm e}$ from
Eqs.~(\ref{eq:FCint}), (\ref{eq:FCext}), (\ref{eq:CDint-ext}) and
(\ref{eq:FGint-ext}) we obtain the equation for $F_n^{\rm i}$\/,
\begin{equation}
\frac{\partial^2 F_n}{\partial t^2} -
   c_{n\rm c}^2\frac{\partial^2 F_n}{\partial z^2} = 0, \quad
c_{n\rm c}^2 = \frac{B^2_0[1+\tanh(ns_0)]}
   {\mu_0[\rho_{\rm i}+\rho_{\rm e}\tanh(ns_0)]},
\label{eq:govF}
\end{equation}
where we have dropped the superscript `i'. Eliminating $D_n^{\rm
i}$\/, $D_n^{\rm e}$ and $G_n^{\rm e}$ from Eqs.~(\ref{eq:GDint}),
(\ref{eq:GDext}), (\ref{eq:CDint-ext}) and (\ref{eq:FGint-ext}) we
obtain the equation for $G_n^{\rm i}$\/,
\begin{equation}
\frac{\partial^2 G_n}{\partial t^2} -
   c_{n\rm s}^2\frac{\partial^2 G_n}{\partial z^2} = 0, \quad
c_{n\rm s}^2 = \frac{B^2_0[1+\tanh(ns_0)]}
   {\mu_0[\rho_{\rm i}\tanh(ns_0)+\rho_{\rm e}]},
\label{eq:govG}
\end{equation}
where we have once again dropped the superscript `i'. It follows
from Eqs.~(\ref{eq:frozen-xi}) and (\ref{eq:xi_int}) that $F_n$ and
$G_n$ have to satisfy the boundary conditions
\begin{equation}
F_n = 0, \quad G_n = 0 \quad \mbox{at} \quad z = \pm L/2.
\label{eq:frozen-FG}
\end{equation}
In Eqs.~(\ref{eq:govF}) and (\ref{eq:govG}) $n = 1$ corresponds to kink modes,
and $n > 1$ to fluting modes.

In the elliptical coordinates the loop axis ($x=y=0$) is defined by
$s = 0$ and $\varphi = \pi/2$\/. It follows from
Eq.~(\ref{eq:xi_int}) that the kink mode described by
Eq.~(\ref{eq:govF}) does not displace the loop axis in the
$s$\/-direction which, at the loop axis, coincides with the
$y$\/-direction. Hence, the loop axis displacement is in the
$x$\/-direction, i.e. this mode is polarised in the direction of the
larger axis of the tube cross-section. The kink mode described by
Eq.~(\ref{eq:govG}) displaces the loop axis in the $s$\/-direction.
It is straightforward to show that it does not displace it in the
$\varphi$\/-direction which, at the loop axis, coincides with the
$x$\/-direction. Hence, the loop axis displacement is in the
$y$\/-direction, i.e. this mode is polarised in the direction of the
smaller axis of the tube cross-section.

When the density is constant, we can use Eqs.~(\ref{eq:govF}) and
(\ref{eq:govG}) with the boundary conditions
Eq.~(\ref{eq:frozen-FG}) to recover the results obtained by
\cite{RUD2003}. Let us look for the eigenmodes and restrict the
analysis to the fundamental modes in the $z$\/-direction. This
implies that we take $F_n$ and $G_n$ proportional to $e^{-i\omega
t}\cos(\pi z/L)$. Then we immediately obtain that the
eigenfrequencies of the boundary value problem defined by
Eq.~(\ref{eq:govF}) and the boundary conditions (\ref{eq:frozen-FG})
are given by
\begin{equation}
\omega_{n\rm c}^2 = \frac{\pi^2 c_{n\rm c}^2}{L^2} =
   \frac{\pi^2 B^2_0[1+\tanh(ns_0)]}
   {\mu_0 L^2[\rho_{\rm i}+\rho_{\rm e}\tanh(ns_0)]},
\quad n = 1,2,\dots,
\label{eq:Feigen}
\end{equation}
and the eigenfrequencies of the boundary value problem defined by
Eq.~(\ref{eq:govG}) and the boundary conditions (\ref{eq:frozen-FG}) are given
by
\begin{equation}
\omega_{n\rm s}^2 = \frac{\pi^2 c_{n\rm s}^2}{L^2} =
   \frac{\pi^2 B^2_0[1+\tanh(ns_0)]}
   {\mu_0 L^2[\rho_{\rm i}\tanh(ns_0)+\rho_{\rm e}]}.
\quad n = 1,2,\dots
\label{eq:Geigen}
\end{equation}
In particular, the squares of eigenfrequencies of the kink modes are given by
\begin{equation}
\omega_{1\rm c}^2 = \frac{\pi^2 B^2_0(a + b)}
   {\mu_0 L^2(a\rho_{\rm i} + b\rho_{\rm e})}, \quad
\omega_{1\rm s}^2 = \frac{\pi^2 B^2_0(a + b)}
   {\mu_0 L^2(b\rho_{\rm i} + a\rho_{\rm e})}.
\label{eq:kink-eigen}
\end{equation}
It is straightforward to see that the eigenfrequencies satisfy
\begin{equation}
\omega_{1\rm c} < \omega_{2\rm c} < \dots < \omega_{2\rm s} < \omega_{1\rm s}.
\label{eq:order}
\end{equation}

\section{Implication on coronal seismology}
\label{sec:seismology}

After \cite{VERetal2004} reported two cases of observations of the
transverse coronal loop oscillations where, in addition to the
fundamental harmonic, the first overtone was also observed,
\cite{ANDetal2005a} suggested observations of this nature could be
used to estimate the scale height in the solar corona.
\cite{ANDetal2005a} assumed that an oscillating loop has a
half-circle shape and a circular cross-section, and it is in the
vertical plane. They also assumed that the atmosphere is isothermal.
In that case, the dependence of the plasma density on $z$ is given
by
\begin{equation}
\rho_{\rm e} = \rho_{\rm f}\exp\left(-\frac L{\pi H}\cos\frac{\pi z}L\right),
\quad \rho_{\rm i} = \zeta\rho_{\rm e},
\label{eq:density}
\end{equation}
where $H$ is the atmospheric scale height, $\rho_{\rm f}$ the plasma density at
the loop foot points outside the loop, and $\zeta > 1$ a constant.
\cite{ANDetal2005a} calculated the ratio of frequencies of the first overtone
and fundamental mode and found that this ratio is a monotonically decreasing
function of the parameter $L/H$\/. Hence, if we know the ratio of frequencies
and $L$\/, we can determine $H$\/. For a recent review of coronal seismology
using kink oscillation overtones see \cite{ANDetal2009}.

A very important question is how robust is this method.
\cite{DYMRUD2006b} and \cite{MORERD2009} have found that the account
of the loop shape can moderately affect the estimates of the
atmospheric scale height. \cite{RUD2007} has shown that the twist of
magnetic field lines in the loop can be safely neglected when
estimating the atmospheric scale height in the corona.
\cite{ROBetal2010} found that the estimates of the atmospheric scale
height obtained using the two-thread model are exactly the same as
those obtained using the model of a monolithic coronal loop with a
circular cross-section of constant radius. Recently \cite{RUD2010}
showed that the account of stationary time independent siphon flows
in coronal loops have little influence {on} the estimates of the
coronal scale height found using the frequency ratio. On the other
hand, \cite{RUDetal2008} and \cite{VERERDJES2008} found that the
account of the loop expansion can strongly affect these estimates.

In this section we study what the effect the elliptic cross-section
has on the estimates of the coronal scale height. As we have already
seen, when a loop has an elliptic cross-section, its kink
oscillations are polarised along the axes of the cross-section. The
kink mode polarised in the direction of the larger axis is described
by Eq.~(\ref{eq:govF}) with $n = 1$, while the kink mode polarised
in the direction of the smaller axis is described by
Eq.~(\ref{eq:govG}) with $n = 1$. Let us consider the solutions to
these equations in the form of eigenmodes and take $F_1$ and $G_1$
proportional to $\exp(-i\omega t)$. Using Eq.~(\ref{eq:density}) we
obtain
\begin{equation}
c_{1\rm c}^2 = \frac{B^2_0(a+b)}{\mu_0\rho_{\rm f}(a\zeta + b)}
   \exp\left(\frac L{\pi H}\cos\frac{\pi z}L\right),
\label{eq:phase-speed-c}
\end{equation}
\begin{equation}
c_{1\rm s}^2 = \frac{B^2_0(a+b)}{\mu_0\rho_{\rm f}(b\zeta + a)}
   \exp\left(\frac L{\pi H}\cos\frac{\pi z}L\right)
\label{eq:phase-speed-s}
\end{equation}
Then, introducing
\begin{equation}
\Omega_{\rm c}^2 = \frac{\mu_0\rho_{\rm f}(a\zeta + b)\omega^2}{B^2_0(a+b)},
   \quad
\Omega_{\rm s}^2 = \frac{\mu_0\rho_{\rm f}(b\zeta + a)\omega^2}{B^2_0(a+b)},
\label{eq:omega-scale}
\end{equation}
we reduce Eqs.~(\ref{eq:govF}) and (\ref{eq:govG}) with $n = 1$ to
\begin{equation}
\frac{d^2 U}{dz^2} + \Omega^2 U
   \exp\left(\frac L{\pi H}\cos\frac{\pi z}L\right) = 0,
\label{eq:govern-scale}
\end{equation}
where either $U = F_1$ and $\Omega = \Omega_{\rm c}$\/, or $U = G_1$
and $\Omega = \Omega_{\rm s}$\/, and $U$ satisfies the boundary
conditions $U = 0$ at $z = \pm L/2$\/. Since
Eq.~(\ref{eq:govern-scale}) does not contain $a$ and $b$\/, the
eigenvalues of the boundary value problem for $U$ are independent of
$a$ and $b$\/. In particular, they are the same as those for a loop
with the circular cross-section. Since
$$
\frac{\Omega_{2\rm c}}{\Omega_{1\rm c}} = \frac{\Omega_2}{\Omega_1}, \qquad
\frac{\Omega_{2\rm s}}{\Omega_{1\rm s}} = \frac{\Omega_2}{\Omega_1},
$$
it follows that we obtain the same estimates of the atmospheric
scale height no matter if we use the observation of the kink
oscillations polarised in the direction of the larger or smaller
axis. The estimates are also independent of $a$ and $b$ and are the
same as those obtained for a loop with the circular cross-section.

\section{Summary and conclusions}
\label{sec:summary}

In this paper we have studied non-axisymmetric oscillations of
straight magnetic loops with a constant elliptic cross-section and
density varying along the loop. We derived the governing equations
for kink and fluting modes in the thin tube approximation. All these
equations are similar to the equation describing kink oscillations
of a straight tube with the circular cross-section. We found that
there are two kink modes, one polarised in the direction of larger
axis of the elliptic cross-section, and the other polarised in the
direction of smaller axis. The frequencies of fundamental mode and
overtones of these two kinds of kink oscillation are different.
However, the ratio of frequencies of the first overtone and the
fundamental mode is the same for both kink oscillations, and it is
independent of the ratio of the ellipse half-axes $a/b$\/. This
result implies that we obtain the same estimates of the atmospheric
scale height no matter if we use the observation of the kink
oscillations polarised in the direction of larger or smaller axis.
The estimates are also the same as those obtained for a loop with
the circular cross-section. This demonstrates that the model shows a
very robust nature when considering a static plasma. However, if the
plasma in the loops is dynamic (i.e. time dependent) then the
ability of the static model to provide accurate estimates may become
questionable (see e.g. \citealp{MORERD2009b}).

\begin{acknowledgements}
The authors thank the Science and Technology Facilities Council
(STFC), UK for the financial support they received.
\end{acknowledgements}

\bibliographystyle{aa}

\end{document}